\documentclass[journal]{IEEEtran}

\usepackage{graphicx,epsfig,subfigure}
\usepackage{amsmath,amssymb,latexsym}
\usepackage{setspace}
\usepackage{longtable}
\usepackage{multicol}
\usepackage{cite,url}
\usepackage{color}

\newtheorem{theorem}{Theorem}
\newtheorem{corollary}{Corollary}
\newtheorem{lemma}{Lemma}

\newtheorem{remark}{Remark}
\newtheorem{fact}{Fact}

 \providecommand{\Av}{\mathbf{A}}
 \providecommand{\Bv}{\mathbf{B}}

 \providecommand{\Fv}{\mathbf{F}}

 \providecommand{\Gv}{\mathbf{G}}
 \providecommand{\Hv}{\mathbf{H}}
 \providecommand{\Iv}{\mathbf{I}}

 \providecommand{\Nv}{\mathbf{N}}
 \providecommand{\Mv}{\mathbf{M}}

 \providecommand{\Sv}{\mathbf{S}}
 
 \providecommand{\Uv}{\mathbf{U}}
 \providecommand{\Vv}{\mathbf{V}}
 \providecommand{\Wv}{\mathbf{W}}

 \providecommand{\Xv}{\mathbf{X}}
 \providecommand{\Yv}{\mathbf{Y}}
 \providecommand{\Zv}{\mathbf{Z}}

 \providecommand{\Cc}{{\mathcal C}}

\providecommand{\Yt}{\widetilde{\mathbf{Y}}}
\providecommand{\yt}{\widetilde{y}}
\providecommand{\Zt}{\widetilde{\mathbf{Z}}}

\providecommand{\Nt}{\widetilde{\mathbf{N}}}

\providecommand{\T}{\intercal}

\begin{document}
\title{New Results on Multiple-Input Multiple-Output Broadcast Channels with Confidential Messages}
\IEEEoverridecommandlockouts

\author{Ruoheng~Liu, Tie~Liu, H.~Vincent~Poor, and Shlomo~Shamai~(Shitz)%
\thanks{This research was supported by the National Science Foundation under Grant
CNS-09-05398, CCF-08-45848 and CCF-09-16867, by the Air Force Office
of Scientific Research under Grant FA9550-08-1-0480, by the European
Commission in the framework of the FP7 Network of Excellence in
Wireless Communications NEWCOM++, and by the Israel Science Foundation. The material in this paper was presented in part at the IEEE International Symposium on Information Theory (ISIT), Austin, TX, June 2010.}%
\thanks{Ruoheng Liu is with Alcatel-Lucent, Murray Hill, NJ 07974, USA (email: ruoheng.liu@alcatel-lucent.com).}%
\thanks{Tie Liu is with the Department of Electrical and Computer Engineering, Texas
A\&M University, College Station, TX 77843, USA (e-mail: tieliu@tamu.edu).}%
\thanks{H. Vincent Poor is with the Department of Electrical Engineering,
Princeton University, Princeton, NJ 08544, USA (e-mail: poor@princeton.edu).}%
\thanks{Shlomo Shamai (Shitz) is with the Department of Electrical Engineering,
Technion-Israel Institute of Technology, Technion City, Haifa 32000,
Israel (e-mail: sshlomo@ee.technion.ac.il).}%
}

\maketitle

\begin{abstract}
This paper presents two new results on multiple-input multiple-output (MIMO) Gaussian broadcast channels with confidential messages. First, the problem of the MIMO Gaussian wiretap channel is revisited. A matrix characterization of the capacity-equivocation region is provided, which extends the previous result on the secrecy capacity of the MIMO Gaussian wiretap channel to the general, possibly imperfect secrecy setting. Next, the problem of MIMO Gaussian broadcast channels with two receivers and three independent messages: a common message intended for both receivers, and two confidential messages each intended for one of the receivers but needing to be kept asymptotically perfectly secret from the other, is considered. A precise characterization of the capacity region is provided, generalizing the previous results which considered only two out of three possible messages.
\end{abstract}


\section{Introduction}\label{sec:INT}
Information-theoretic security has been a very active area of research recently. (See \cite{LPS-M09} and \cite{LT-M10} for overviews of recent progress in this field.) In particular, significant progress has been made in understanding the fundamental limits of multiple-input multiple-output (MIMO) secret communication. More specifically, the secrecy capacity of the MIMO Gaussian wiretap channel was characterized in \cite{KW-IT10a,KW-IT10b,OH-ITS,LS-IT09,BLPS-EURASIP09}. The works \cite{LP-IT09} and \cite{LLPS-IT10} considered the problem of MIMO Gaussian broadcast channels with two confidential messages, each intended for one receiver but needing to be kept asymptotically perfectly secret from the other, and provided a precise characterization of the capacity region. The capacity region of the MIMO Gaussian broadcast channel with two receivers and two independent messages, a common message intended for both receivers and a confidential message intended for one of the receivers but needing to be kept asymptotically perfectly secret from the other, was characterized in \cite{LLL-IT10}.

This paper presents two new results on MIMO Gaussian broadcast channels with confidential messages\footnote{The main results of this paper were initially posted on the arXiv website in January 2010 \cite{LLPS-A10} and were subsequently reported at the 2010 IEEE International Symposium on Information Theory \cite{LLPS-ISIT10a,LLPS-ISIT10b}. Similar results were independently reported by Ekrem and Ulukus in \cite{EU-ITSa} and \cite{EU-ITSb}.
}:
\begin{itemize}
\item[1)] The problem of the MIMO Gaussian wiretap channel is revisited. A matrix characterization of the \emph{capacity-equivocation} region is provided, which extends the result of \cite{LS-IT09} on the secrecy capacity of the MIMO Gaussian wiretap channel to the general, possibly imperfect secrecy setting.
\item[2)] The problem of MIMO Gaussian broadcast channels with two receivers and \emph{three} independent messages, a common message intended for both receivers, and two mutually confidential messages each intended for one of the receivers but needing to be kept asymptotically perfectly secret from the other, is considered. A precise characterization of the capacity region is provided, generalizing the results of \cite{LLPS-IT10} and \cite{LLL-IT10} which considered only two out of three possible messages.
\end{itemize}

\emph{Notation}. Vectors and matrices are written in bold letters. All vectors by default are column vectors. The identity matrices are denoted by $\Iv$, where a subscript may be used to indicate the size of the matrix to avoid possible confusion. The transpose of a matrix $\Av$ is denoted by $\Av^{\T}$, and the trace of a square matrix $\Av$ is denoted by $\mathrm{Tr}(\Av)$. Finally, we write $\Av \preceq \Bv$ (or, equivalently, $\Bv \succeq \Av$) whenever $\Bv-\Av$ is positive semidefinite.

\begin{figure*}[t!]
\centering
\subfigure[Rate-equivocation setting]{\includegraphics[width=0.65\linewidth,draft=false]{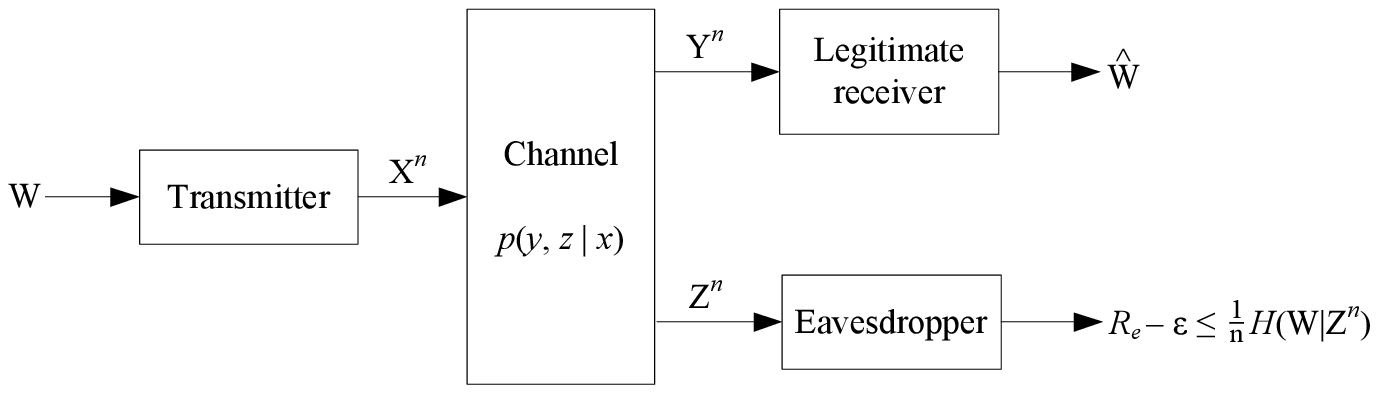}}\vspace{10pt}\\
\subfigure[Simultaneous private-confidential
communication]{\includegraphics[width=0.65\linewidth,draft=false]{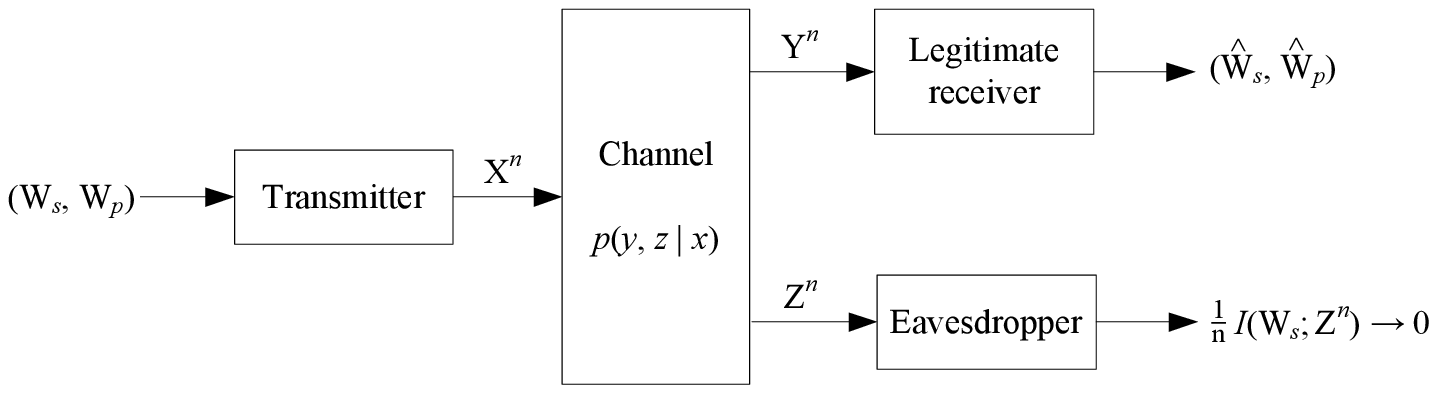}}
\caption{Wiretap channel.} \label{fig:WTC}
\end{figure*}

\section{The Capacity-Equivocation Region of the MIMO Gaussian Wiretap Channel}\label{sec:ce}
\subsection{Channel Model}\label{sec:ce-cm}
Consider a MIMO Gaussian broadcast channel with two receivers, one of which is a legitimate receiver and the other is an eavesdropper. The received signals at time index $m$ are given by
\begin{equation}
\begin{array}{rll}
\Yv[m] &=& \Hv_r\mathbf{X}[m]+\Wv_r[m]\\
\Zv[m] &=& \Hv_e\mathbf{X}[m]+\Wv_e[m]
\end{array}
\label{eq:Ch}
\end{equation}
where $\Hv_r$ and $\Hv_e$ are (real) channel matrices at the
legitimate receiver and the eavesdropper respectively, and
$\{\Wv_r[m]\}_m$ and $\{\Wv_e[m]\}_m$ are independent and
identically distributed (i.i.d.) additive vector Gaussian noise processes
with zero means and \emph{identity} covariance matrices.

The transmitter has a single message $W$, which is uniformly distributed
over $\{1,\ldots,2^{nR}\}$ where $R$ is the \emph{rate} of
communication. The goal of communication is to deliver $W$ reliably
to the legitimate receiver while keeping it information-theoretically secure from
the eavesdropper. Following the classical work \cite{Wyn-BSTJ75,CK-IT78},
for every $\epsilon>0$ it is required that
\begin{equation}
\frac{1}{n}H(W|\Zv^n) \geq R_e-\epsilon \label{eq:eqv}
\end{equation}
for sufficiently large $n$, where $n$ is the block length of
communication, $\Zv^n:=(\Zv[1],\ldots,\Zv[n])$, and $R_e$ represents the
predetermined level of security of message $W$ at the eavesdropper known as \emph{equivocation}. The
\emph{capacity-equivocation} region is the set of rate-equivocation
pairs $(R,R_e)$ that can be achieved by \emph{any} coding scheme. In the literature, this communication scenario is usually known as the rate-equivocation setting of the MIMO Gaussian \emph{wiretap} channel; see Fig.~\ref{fig:WTC}(a) for an illustration.

Csisz\'{a}r and K\"{o}rner\cite{CK-IT78} studied the rate-equivocation setting of a general discrete memoryless wiretap channel. A single-letter expression for the capacity-equivocation region was derived \cite[Theorem~1]{CK-IT78}, which can be written as the set of nonnegative
rate-equivocation pairs $(R,R_e)$ satisfying
\begin{equation}
\begin{array}{rcl}
R_e &\le & \min\{R,I(V;Y|U) - I(V;Z|U)\}\\
R &\le & I(V;Y) \label{eq:CK}
\end{array}
\end{equation}
for some $p(u,v,x,y,z)=p(u)p(v|u)p(x|v)p(y,z|x)$. Here, $p(y,z|x)$ is the transition probability of the discrete memoryless wiretap channel, and $U$ and $V$ are two \emph{auxiliary} random variables. In theory, a computable expression for the capacity-equivocation region can be obtained by evaluating the single-letter expression \eqref{eq:CK} for the MIMO Gaussian wiretap channel \eqref{eq:Ch}. However, such an evaluation is generally difficult due to the presence of the auxiliary random variables $U$ and $V$.

Several recent works \cite{KW-IT10a,KW-IT10b,OH-ITS,LS-IT09,BLPS-EURASIP09} studied the special case where the equivocation $R_e$ is set to equal the communication rate $R$. In this case, the secrecy constraint \eqref{eq:eqv} can be equivalently written as
\begin{equation}
\frac{1}{n}I(W;\Zv^n) \leq \epsilon \label{eq:eqv1}
\end{equation}
i.e., message $W$ needs to be asymptotically \emph{perfectly} secure
from the eavesdropper. Under the asymptotic perfect secrecy constraint
\eqref{eq:eqv}, the maximum rate of communication is called the
\emph{secrecy} capacity. For the MIMO Gaussian
wiretap channel \eqref{eq:Ch}, a matrix characterization of the
secrecy capacity was obtained in \cite{KW-IT10a,KW-IT10b,OH-ITS} under an average total power constraint and in \cite{LS-IT09} and\cite{BLPS-EURASIP09} under a more general \emph{matrix} power constraint. Similar matrix characterizations of the capacity-equivocation region, however, were \emph{unknown}.

\subsection{Main Results}
The main result of this section is a matrix characterization of the capacity-equivocation region of the MIMO Gaussian wiretap channel. More specifically, consider the MIMO Gaussian wiretap channel \eqref{eq:Ch} under the matrix power constraint
\begin{equation}
\frac{1}{n}\sum_{m=1}^{n}(\mathbf{X}[m]\mathbf{X}^{\T}[m]) \preceq
\mathbf{S} \label{eq:MPC}
\end{equation}
where $\Sv$ is a positive semidefinite matrix. Let
\begin{equation}
C(\Sv,\Hv_r)=\frac{1}{2}\log\left|\Iv+\Hv_r\Sv\Hv_r^{\T}\right|
\label{eq:C}
\end{equation}
be the Shannon capacity of a MIMO Gaussian point-to-point channel
with channel matrix $\Hv_r$ and under the matrix power constraint
\eqref{eq:MPC}, and let
\begin{equation}
C_s(\Sv,\Hv_r,\Hv_e) =\max_{0\preceq\Bv\preceq
\Sv}\frac{1}{2}\log\left|\frac{\Iv+\Hv_r\Bv\Hv_r^{\T}}{\Iv+\Hv_e
\Bv\Hv_e^{\T}}\right|\label{eq:Cs}
\end{equation}
be the secrecy capacity of a MIMO Gaussian wiretap channel with legitimate
receiver and eavesdropper channel matrices $\Hv_r$ and $\Hv_e$
respectively and under the matrix power constraint \eqref{eq:MPC}
\cite{LS-IT09,BLPS-EURASIP09}. We then have the following result.

\begin{theorem} \label{thm:ce}
The capacity-equivocation region of the MIMO Gaussian wiretap channel \eqref{eq:Ch} under the matrix power constraint
\eqref{eq:MPC} is given by the set of nonnegative rate-equivocation
pairs $(R,R_e)$ satisfying
\begin{equation}
\begin{array}{rcl}
R_e & \le & \min\{R,C_s(\Sv,\Hv_r,\Hv_e)\}\\
R & \le & C(\Sv,\Hv_r)
\end{array}
\end{equation}
where $C(\Sv,\Hv_r)$ and $C_s(\Sv,\Hv_r,\Hv_e)$ are defined as in
\eqref{eq:C} and \eqref{eq:Cs}, respectively.
\end{theorem}

\begin{figure}[t!]
\centering
\subfigure[Capacity-equivocation region]{
\includegraphics[width=0.9\linewidth,draft=false]{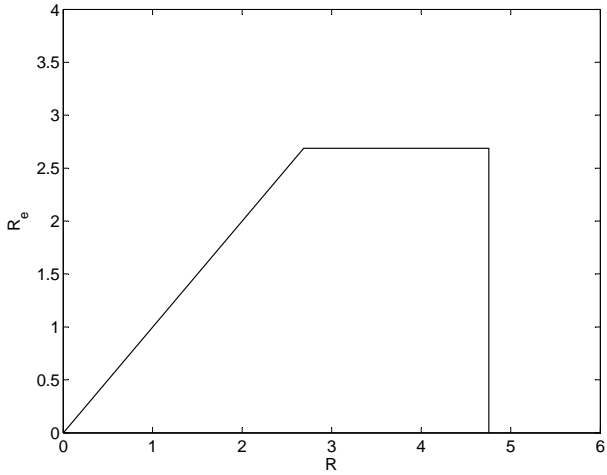}}\\[5mm]
\subfigure[Private-confidential message capacity
region]{
\includegraphics[width=0.9\linewidth,draft=false]{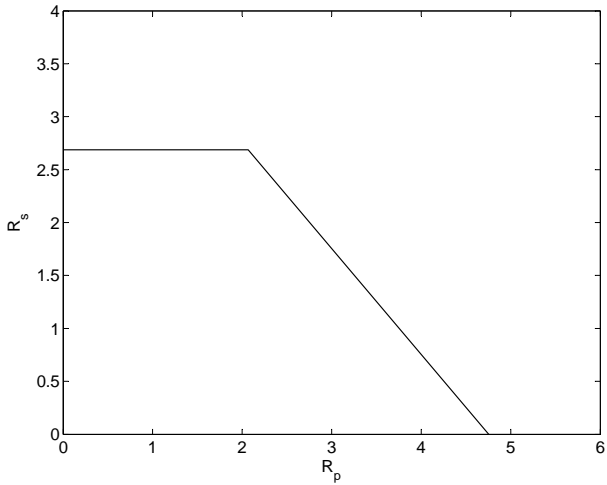}}
\caption{MIMO Gaussian wiretap channel under matrix power constraint.} \label{fig:ce}
\end{figure}

Fig.~\ref{fig:ce}(a) illustrates the capacity-equivocation region of a MIMO Gaussian wiretap channel with channel matrices
\begin{align*}
\Hv_r &=\left(\begin{matrix} 1.8&  2.0 \\ 1.0 &  3.0
\end{matrix}\right) \quad \mbox{and} \quad
\Hv_e =\left(\begin{matrix} 3.3 &  1.3 \\
2.0 & -1.5
\end{matrix}\right)
\end{align*}
(which yields a \emph{nondegraded} wiretap channel) and matrix power constraint
\begin{align*}
\Sv=\left(\begin{matrix} 5.0 & 1.25 \\
1.25 & 10.0 \end{matrix}\right).
\end{align*}

The capacity-equivocation region of the MIMO Gaussian wiretap channel under an average total power constraint is summarized in the following corollary. The result is a direct consequence of Theorem~\ref{thm:ce} and \cite[Lemma~1]{WSS-IT06}.

\begin{corollary}
The capacity-equivocation region of the MIMO Gaussian wiretap
channel \eqref{eq:Ch} under the average total power constraint
\begin{equation}
\frac{1}{n}\sum_{m=1}^{n}\left(\mathbf{X}[m]^{\T}\mathbf{X}[m]\right)
\leq P \label{eq:ATPC}
\end{equation}
is given by the set of nonnegative rate-equivocation
pairs $(R,R_e)$ satisfying
\begin{equation}
\begin{array}{rcl}
R_e &\le &\min\{R,C_s(\Sv,\Hv_r,\Hv_e)\}\\
R &\le & C(\Sv,\Hv_r)
\end{array}
\end{equation}
for some $\Sv \succeq 0$, $\mathrm{Tr}(\Sv) \leq P$.
\end{corollary}

\subsection{Proof of the Main Results}
Next, we prove Theorem~\ref{thm:ce}. As mentioned previously, directly evaluating the single-letter expression \eqref{eq:CK} for the MIMO Gaussian wiretap channel \eqref{eq:Ch} is difficult due to the presence of the auxiliary random variables. We thus resort to an \emph{indirect} approach that connects the rate-equivocation setting of a MIMO Gaussian wiretap channel to the problem of simultaneously communicating private and confidential messages.

The problem of simultaneously communicating private and confidential messages over a discrete memoryless wiretap channel is illustrated in Fig.~\ref{fig:WTC}(b). Here, the transmitter has a private message $W_p$, which is uniformly distributed over $\{1,\ldots,2^{nR_p}\}$, and a confidential message
$W_s$, which is uniformly distributed over $\{1,\ldots,2^{nR_s}\}$.
The confidential message $W_s$ is intended for the legitimate
receiver but needs to be kept asymptotically \emph{perfectly} secret from
the eavesdropper. That is, for any $\epsilon>0$ it is required that
\begin{align}
\frac{1}{n}I(W_s;Z^n) \leq \epsilon \label{eq:eqv2}
\end{align}
for sufficiently large block length $n$. The private message $W_p$ is also
intended for the legitimate receiver, but is \emph{not} subject to
any secrecy constraint. The \emph{private-confidential} message capacity region is the set of private-confidential rate pairs $(R_p,R_s)$ that can be achieved by
\emph{any} coding scheme.

The following lemma provides a single-letter characterization of the private-confidential message capacity region of the discrete memoryless wiretap channel.

\begin{figure*}[t!]
 \centerline{\includegraphics[width=0.7\linewidth,draft=false]{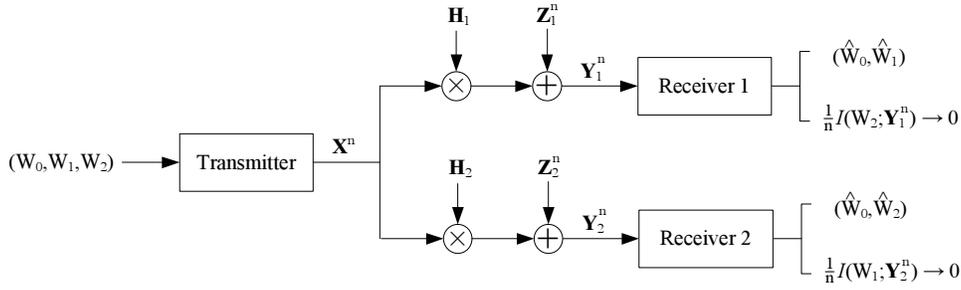}}
 \caption{MIMO Gaussian broadcast channel with common and confidential messages.}
 \label{fig:chm}
\end{figure*}

\begin{lemma}\label{lemma:DMC}
The private-confidential message capacity region of the discrete memoryless wiretap channel $p(y,z|x)$ is given by the set of nonnegative private-confidential rate pairs $(R_p, R_s)$ satisfying
\begin{equation}
\begin{array}{rcl}
R_s &\le&  I(V;Y|U) - I(V;Z|U)\\
R_s+R_p &\le & I(V;Y)
\end{array}
\label{eq:SR-DMC}
\end{equation}
for some $p(u,v,x,y,z)=p(u)p(v|u)p(x|v)p(y,z|x)$, where $U$ and $V$
are auxiliary random variables.
\end{lemma}

The achievability part of the lemma can be proved by considering a coding scheme that combines superposition coding, random binning, and rate splitting. In particular, part of the private message will be used in the binning scheme to protect the confidential message against the eavesdropper. The converse proof follows standard information-theoretic argument. The details of the proof are deferred to Appendix~\ref{app:L1}.

A simple inspection of the capacity-equivocation region \eqref{eq:CK} and the private-confidential message capacity region \eqref{eq:SR-DMC} reveals the following interesting fact:

\begin{fact}\label{fact}
A nonnegative rate pair $(R,R_e)=(R_p+R_s,R_s)$ is an achievable rate-equivocation pair for a discrete memoryless wiretap channel if and only if $(R_p,R_s)$ is an achievable private-confidential rate pair for the same channel.
\end{fact}

The ``if" part of the fact is easy to verify: Simply use the \emph{same} code for both communication scenarios and view $(W_p,W_s)$ as the single message $W$ for the rate-equivocation setting. Note that
\begin{eqnarray*}
\frac{1}{n}H(W|Z^n) &=& \frac{1}{n}H(W_p,W_s|Z^n)\\
& \geq & \frac{1}{n}H(W_s|Z^n)\\
& \geq & R_s-\epsilon\\
& = & R_e -\epsilon.
\end{eqnarray*}
Thus, the same code satisfying the secrecy constraint \eqref{eq:eqv2} for simultaneous private-confidential communication also satisfies the secrecy constraint \eqref{eq:eqv} for the rate-equivocation setting. The ``only if" part of the fact comes as a mild surprise, as in the rate-equivocation setting which part of message is secure does not need to be specified \emph{a priori} and may even depend on the realization of the channel noise. We note here that the above interesting fact was first mentioned in \cite[pp.~411--412]{CK-B82} without proof.

In light of Fact~\ref{fact}, next we first establish a matrix characterization of the private-confidential message capacity region using the existing matrix characterization \cite{LS-IT09,BLPS-EURASIP09} on the secrecy capacity of the MIMO Gaussian wiretap channel. The result will then be mapped to the rate-equivocation setting using the aforementioned equivalence between these two communication scenarios.

\begin{lemma}\label{lemma:GMW}
The private-confidential message capacity region of the MIMO Gaussian wiretap channel \eqref{eq:Ch} under the matrix power constraint \eqref{eq:MPC} is given by the set of nonnegative private-confidential rate pairs $(R_p,R_s)$ satisfying
\begin{equation}
\begin{array}{rcl}
R_s & \le & C_s(\Sv,\Hv_r,\Hv_e)\\
R_s+R_p & \le & C(\Sv,\Hv_r).
\end{array}
\label{eq:PC}
\end{equation}
\end{lemma}

\begin{IEEEproof}
Let $\Bv^*$ be an optimal solution to the optimization problem on
the right-hand side of \eqref{eq:Cs}. Then, the achievability of the
private-confidential rate region \eqref{eq:PC} follows from that of
\eqref{eq:SR-DMC} by setting $\Vv=\Xv=\Uv+\Gv$, where $\Uv$ and $\Gv$ denote two
independent Gaussian vectors with zero means and covariance matrices
$\Sv-\Bv^*$ and $\Bv^*$, respectively.

The fact that $R_s \le C_s(\Sv,\Hv_r,\Hv_e)$ for any achievable confidential rate $R_s$ follows from the secrecy capacity result of \cite{LS-IT09} and \cite{BLPS-EURASIP09} on the MIMO Gaussian wiretap channel under a matrix power constraint, by ignoring the private message $W_p$. The fact that $R_s+R_p \le C(\Sv,\Hv_r)$ for any achievable private-confidential rate pair $(R_p,R_s)$ follows from the well-known capacity result on the MIMO Gaussian point-to-point channel under a matrix power constraint, by viewing $(W_p,W_s)$ as a single message and ignoring the asymptotic perfect secrecy constraint \eqref{eq:eqv2} on the confidential message $W_s$.
\end{IEEEproof}

\begin{remark}
It is particularly worth mentioning the corner point
$(R_p,R_s)$ of the private-confidential message capacity region \eqref{eq:PC} as given by
$$(R_p,R_s) = \left(C(\Sv,\Hv_r)-C_s(\Sv,\Hv_r,\Hv_e),C_s(\Sv,\Hv_r,\Hv_e)\right).$$
Here, under the matrix power constraint, both messages $W_s$ and $(W_p,W_s)$, viewed as a single
private message, can transmit \emph{simultaneously} at their
respective \emph{maximum} rates. In particular, transmitting an additional private message $W_p$
does \emph{not} incur any rate loss for communicating the confidential message
$W_s$.
\end{remark}

Now, Theorem~\ref{thm:ce} follows immediately from Lemma~\ref{lemma:GMW} and a Fourier-Motzkin elimination with $R=R_p+R_s$ and $R_e=R_s$. For comparison, the private-confidential message capacity region of the same MIMO Gaussian wiretap channel as used for Fig.~\ref{fig:ce}(a) is illustrated in Fig.~\ref{fig:ce}(b).

\section{MIMO Gaussian Broadcast Channels with Common and Confidential Messages}
\subsection{Channel Model}
Consider a two-receiver MIMO Gaussian broadcast channel. The transmitter is equipped with $t$ transmit antennas, and receiver~$k$, $k=1,2$, is equipped with $r_k$ receive
antennas. A discrete-time sample of the channel at time $m$ can be
written as
\begin{equation}
\Yv_{k}[m] = \Hv_k\mathbf{X}[m]+\Zv_k[m], \quad k=1,2 \label{eq:Ch1}
\end{equation}
where $\Hv_k$ are the (real) channel matrices of size $r_{k} \times t$,
and $\{\Zv_k[m]\}_m$ are i.i.d. additive vector Gaussian noise processes with zero means and
identity covariance matrices.\footnote{The channel model is the same as that in Section~\ref{sec:ce-cm}. However, different notation is used here for the convenience of presentation.}

As illustrated in Fig.~\ref{fig:chm}, the transmitter has a common message $W_0$ and two
independent confidential messages $W_1$ and $W_2$. The common message $W_0$ is intended for both receivers. The confidential message $W_k$ is intended for receiver $k$ but needs to be kept asymptotically perfectly secret from the other receiver. Mathematically, for every $\epsilon>0$ we must have
\begin{equation}
\frac{1}{n}I(W_1;\Yv_{2}^n) \leq \epsilon \quad \mbox{and} \quad
\frac{1}{n}I(W_2;\Yv_{1}^n) \leq \epsilon \label{eq:PS-1}
\end{equation}
for sufficiently large block length $n$. Our goal here is to
characterize the entire capacity region $\Cc(\Hv_1,\Hv_2,\Sv)=\{(R_0,R_1,R_2)\}$ that can be achieved by
any coding scheme, where $R_0$, $R_1$ and $R_2$ are the
communication rates corresponding to the common message $W_0$ and the
confidential messages $W_1$ and $W_2$, respectively.

With both confidential messages $W_1$ and $W_2$ but \emph{without}
the common message $W_0$, the problem was studied in \cite{LP-IT09}
for the multiple-input single-output (MISO) case and in
\cite{LLPS-IT10} for general MIMO case. Rather surprisingly, it
was shown in \cite{LLPS-IT10} that, under a matrix power
constraint both confidential messages can be
\emph{simultaneously} communicated at their respected maximum rates. With the common message $W_0$ and only \emph{one} confidential message ($W_1$ or $W_2$), the capacity region of the MIMO Gaussian wiretap channel was
characterized in \cite{LLL-IT10} using a channel-enhancement
approach \cite{WSS-IT06} and an extremal entropy inequality of
Weingarten {\it et al.} \cite{WLSSV-IT09}.

\subsection{Main Results}
The main result of this section is a precise characterization of the capacity region of the MIMO Gaussian
broadcast channel with a more complete message set that includes a common message $W_0$ and two independent confidential messages $W_1$ and $W_2$.

\begin{theorem}\label{thm:GMBC}
The capacity region $\Cc(\Hv_1,\Hv_2,\Sv)$ of the MIMO Gaussian broadcast channel
\eqref{eq:Ch1} with a common message $W_0$ and two confidential messages $W_1$ and $W_2$ under
the matrix power constraint \eqref{eq:MPC} is given by the set of nonnegative rate triples $(R_0,R_1,R_2)$ such that
\begin{equation}
\begin{array}{rcl}
R_0 & \le & \min\left\{\frac{1}{2}\log
\left|\frac{\Hv_1\Sv\Hv_1^{\T}+\Iv_{r1}}{\Hv_1(\Sv-\Bv_0)\Hv_1^{\T}+\Iv_{r1}}\right|,\right.\\
& & \hspace{40pt} \left.\frac{1}{2}\log\left|\frac{\Hv_2\Sv\Hv_2^{\T}+\Iv_{r2}}{\Hv_2(\Sv-\Bv_0)\Hv_2^{\T}+\Iv_{r2}}\right|
\right\}\\
R_1 & \le &
\frac{1}{2}\log\left|\Iv_{r_1}+\Hv_1\Bv_1\Hv_1^{\T}\right|-\\
& & \hspace{40pt} \frac{1}{2}\log\left|\Iv_{r_2}+\Hv_2
\Bv_1\Hv_2^{\T}\right|\\
\qquad R_2 & \le &
\frac{1}{2}\log\left|\frac{\Iv_{r_2}+\Hv_2(\Sv-\Bv_0)\Hv_2^{\T}}{\Iv_{r_2}+\Hv_2\Bv_1\Hv_2^{\T}}\right|-\\
& & \hspace{40pt} \frac{1}{2}\log\left|\frac{\Iv_{r_1}+\Hv_1(\Sv-\Bv_0)\Hv_1^{\T}}{\Iv_{r_1}+\Hv_1\Bv_1\Hv_1^{\T}}\right|
\end{array}
\label{eq:SCR}
\end{equation}
for some $\Bv_0 \succeq 0$, $\Bv_1 \succeq 0$ and $\Bv_0+\Bv_1
\preceq \Sv$.
\end{theorem}

\begin{figure}[t]
\centering
\subfigure[Capacity region $\Cc(\Hv_1,\Hv_2,\Sv)$]{
\includegraphics[width=0.9\linewidth,draft=false]{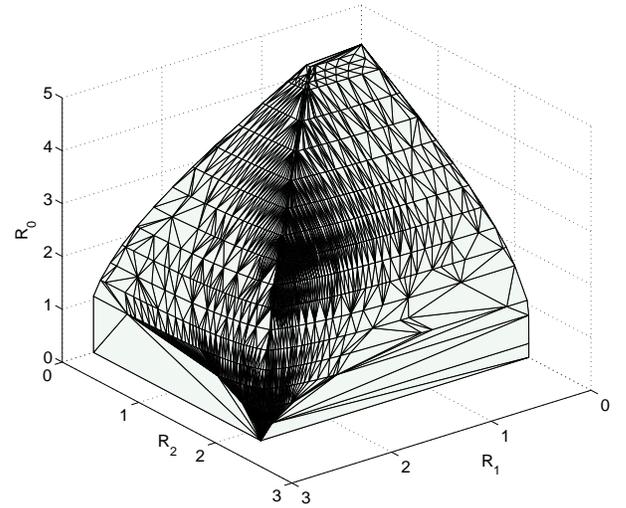}}\\[5mm]
\subfigure[$(R_1,R_2)$-cross sections]{
\includegraphics[width=0.9\linewidth,draft=false]{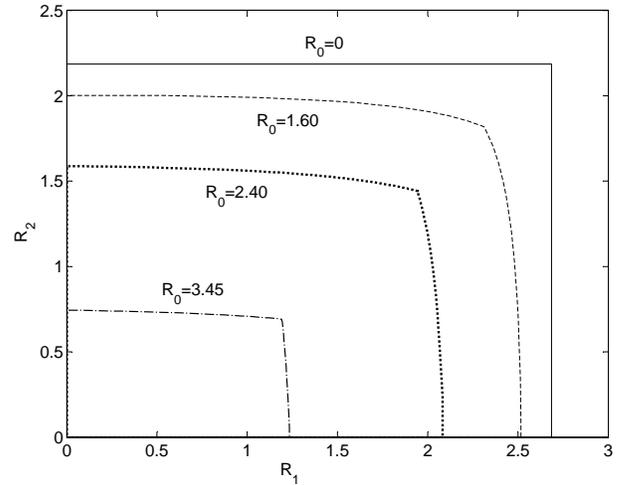}}
\caption{MIMO Gaussian broadcast channel with common and confidential messages.}
\label{fig:ccm2}
\end{figure}

\begin{remark}
By setting $\Bv_0=0$ we can recover the result of \cite[Theorem~1]{LLPS-IT10} that includes both confidential messages $W_1$ and $W_2$ but without the common message $W_0$. Similar to \cite[Theorem~1]{LLPS-IT10}, for any given $\Bv_0$ the upper bounds on $R_1$ and $R_2$ can be simultaneously maximized by a same $\Bv_1$. In fact, the upper bounds on $R_1$ and $R_2$ in \eqref{eq:SCR} are fully symmetric with respect to $\Hv_1$ and $\Hv_2$, even though it is not immediately evident from the expressions themselves.
\end{remark}

\begin{remark}
By setting $\Bv_0=\Sv-\Bv_1$ we can recover the result of \cite[Theorem~1]{LLL-IT10} that includes the common message $W_0$ and the confidential message $W_1$ but without the other confidential message $W_2$.
\end{remark}

Fig.~\ref{fig:ccm2}(a) illustrates the capacity region $\Cc(\Hv_1,\Hv_2,\Sv)$ for the channel matrices and the matrix
power constraint as given by
\begin{align*}
\Hv_1 &=\left(\begin{matrix} 1.8&  2.0 \\ 1.0 &  3.0
\end{matrix}\right),~
\Hv_2 =\left(\begin{matrix} 3.3 &  1.3 \\
2.0 & -1.5
\end{matrix}\right)\\
\text{and} \qquad &\qquad
\Sv=\left(\begin{matrix} 5.0 & 1.25 \\ 1.25 & 10.0 \end{matrix}\right).
\end{align*}
(The channel parameters are the same as those used for Fig.~\ref{fig:ce}.) In
Fig.~\ref{fig:ccm2}(b), we have also plotted the $(R_1,R_2)$-cross section of
$\Cc(\Hv_1,\Hv_2,\Sv)$ for several given values of $R_0$. Note that when $R_0=0$, the
$(R_1,R_2)$-cross section is \emph{rectangular}, implying that under a matrix power
constraint, both confidential messages $W_1$ and $W_2$ can be simultaneously transmitted
at their respective maximum rates \cite{LLPS-IT10}. For $R_0>0$, however, the
$(R_1,R_2)$-cross sections are generally non-rectangular as different boundary points on
the same cross section may correspond to \emph{different} choice of $\Bv_0$.

%

The capacity region under an average total power constraint is summarized in the
following corollary. The result is a direct consequence of Theorem~\ref{thm:GMBC} and
\cite[Lemma~1]{WSS-IT06}.

\begin{corollary}
The capacity region $\Cc(\Hv_1,\Hv_2,P)$ of the MIMO Gaussian broadcast channel
\eqref{eq:Ch1} with a common message $W_0$ and two confidential messages $W_1$ and $W_2$ under
the average total power constraint \eqref{eq:ATPC} is given by
\begin{equation}
\Cc(\Hv_1,\Hv_2,P) = \bigcup_{\Sv \succeq 0, \; \mathrm{Tr}(\Sv) \leq P} \Cc(\Hv_1,\Hv_2,\Sv).
\end{equation}
\end{corollary}

\subsection{Proof of the Main Results}
Next, we prove Theorem~\ref{thm:GMBC}. Following \cite{WSS-IT06}, we shall focus on the canonical case in which the channel matrices $\Hv_1$ and $\Hv_2$ are square and invertible and the matrix power constraint $\Sv$ is strictly positive definite. In this case, multiplying both sides of \eqref{eq:Ch1} by $\Hv_k^{-1}$, the MIMO Gaussian broadcast channel \eqref{eq:Ch1} can be equivalently written as
\begin{equation}
\Yv_k[m] =\Xv_k[m]+\Zv_k[m], \quad k=1,2 \label{eq:CH-A}
\end{equation}
where $\{\Zv_k[m]\}_m$ are i.i.d. additive vector Gaussian noise
processes with zero means and covariance matrices
$\Nv_k=\Hv_k^{-1}\Hv_{k}^{-\T}$. Similarly, the rate region \eqref{eq:SCR} can be equivalently written as
\begin{equation}
\begin{array}{rcl}
R_0 & \le & \min\left\{\frac{1}{2}\log
\left|\frac{\Sv+\Nv_1}{(\Sv-\Bv_0)+\Nv_1}\right|,\frac{1}{2}\log
\left|\frac{\Sv+\Nv_2}{(\Sv-\Bv_0)+\Nv_2}\right|
\right\}\\
R_1 & \le &
\frac{1}{2}\log\left|\frac{\Bv_1+\Nv_1}{\Nv_1}\right|-\frac{1}{2}\log\left|\frac{
\Bv_1+\Nv_2}{\Nv_2}\right|\\
R_2 & \le & \frac{1}{2}\log\left|\frac{
(\Sv-\Bv_0)+\Nv_2}{\Bv_1+\Nv_2}\right|-
\frac{1}{2}\log\left|\frac{(\Sv-\Bv_0)+\Nv_1}{\Bv_1+\Nv_1}\right|.
\end{array}
\label{eq:SCR-A}
\end{equation}

Next, we show that the rate region \eqref{eq:SCR-A} over all possible $\Bv_0 \succeq 0$, $\Bv_1 \succeq 0$ and $\Bv_0+\Bv_1 \preceq \Sv$ gives the capacity region $\Cc(\Hv_1,\Hv_2,\Sv)$ for the canonical MIMO Gaussian broadcast channel \eqref{eq:CH-A}. Extensions to the general model \eqref{eq:Ch1} follow from the well-known limiting argument \cite{WSS-IT06,LS-IT09,LLL-IT10} and hence are omitted from the paper.

To prove the achievability of the rate region \eqref{eq:SCR-A}, recall that the problem of a two-receiver discrete memoryless broadcast channel with a common message and two confidential common messages was
studied in \cite{XCC-IT09}. There, a single-letter expression for an achievable rate region was established, which is given by the set of rate triples $(R_0,R_1,R_2)$ such that
\begin{equation}
\begin{array}{rcl}
R_0 & \le & \min[I(\Uv; \Yv_1),I(\Uv, \Yv_2)]\\
R_1 & \le & I(\Vv_1;\Yv_1|\Uv)-I(\Vv_1;\Vv_2,\Yv_2|\Uv)\\
R_2 & \le & I(\Vv_2;\Yv_2|\Uv)-I(\Vv_2;\Vv_1,\Yv_1|\Uv) \label{eq:ASRR-LP}
\end{array}
\end{equation}
where $\Uv$, $\Vv_1$ and $\Vv_2$ are auxiliary random variables satisfying the Markov
relation $(\Uv,\Vv_1,\Vv_2)\rightarrow\Xv\rightarrow(\Yv_1,\Yv_2)$. The proposed coding
scheme is a natural combination of double binning \cite{LMSY-IT08} and superposition
coding. Thus, the achievability of the rate region \eqref{eq:SCR-A} follows from that of
(\ref{eq:ASRR-LP}) by setting $\Vv_1 =\Uv_1+\Fv \Uv_2$, $\Vv_2 =\Uv_2$, and $\Xv
=\Uv+\Uv_1+\Uv_2$ where $\Uv$, $\Uv_1$ and $\Uv_2$ are three independent Gaussian vectors
with zero means and covariance matrices $\Bv_0$, $\Bv_1$ and $\Sv-\Bv_0-\Bv_1$
respectively, and $$\Fv :=\Bv\Hv_1^{\T}(\Iv_{r_1}+\Hv_1\Bv\Hv_1^{\T})^{-1}\Hv_1.$$

To show that the rate region \eqref{eq:SCR-A} over all possible $\Bv_0 \succeq 0$, $\Bv_1 \succeq 0$ and $\Bv_0+\Bv_1 \preceq \Sv$ is indeed the capacity region, we shall consider proof by contradiction and resort to a channel-enhancement argument akin to that in \cite{LLL-IT10}.

More specifically, assume that $(R_0^{\dag},R_1^{\dag},R_2^{\dag})$ is an \emph{achievable} rate triple that lies \emph{outside} the rate region \eqref{eq:SCR-A} for any given $\Bv_0 \succeq 0$, $\Bv_1 \succeq 0$ and $\Bv_0+\Bv_1 \preceq \Sv$. Since
$(R_0^{\dag},R_1^{\dag},R_2^{\dag})$ is achievable, we can bound
$R_0^{\dag}$ by
\begin{align*}
R_0^{\dag} &\le \min \left(
\frac{1}{2}\log\left|\frac{\Sv+\Nv_1}{\Nv_1}\right|,\;
\frac{1}{2}\log\left|\frac{\Sv+\Nv_2}{\Nv_2}\right|\right)= R_0^{\rm max}.
\end{align*}
Moreover, if $R_1^{\dag}=R_2^{\dag}=0$, then $R_0^{\rm max}$ can be
achieved by setting $\Bv_0=\Sv$ and $\Bv_1=0$ in \eqref{eq:SCR-A}.
Thus, by the assumption that $(R_0^{\dag},R_1^{\dag},R_2^{\dag})$ is {outside} the rate region \eqref{eq:SCR-A} for any given $\Bv_0 \succeq 0$, $\Bv_1 \succeq 0$ and $\Bv_0+\Bv_1 \preceq \Sv$, we can always find $\lambda_1\ge 0$ and $\lambda_2 \ge 0$ such that
\begin{align}
\lambda_1 R_1^{\dag}+\lambda_2 R_2^{\dag}=\lambda_1 R_1^{\star}+\lambda_2
R_2^{\star}+\rho
\end{align}
for some $\rho > 0$, where $\lambda_1 R_1^{\star}+\lambda_2 R_2^{\star}$ is
given by
\begin{equation}
\begin{array}{rl}
\max_{(\Bv_0,\Bv_1)} & \lambda_1 f_1(\Bv_1)+\lambda_2 f_2(\Bv_0,\Bv_1)\\
\text{subject to} & f_0(\Bv_0) \ge R_0^{\dag}\\
                          &  \Bv_0 \succeq 0\\
                          &  \Bv_1 \succeq 0\\
                          & \Bv_0+\Bv_1 \preceq \Sv.
\end{array}
\label{eq:op-A}
\end{equation}
Here, the functions $f_0$, $f_1$ and $f_2$ are defined as
\begin{align}
f_0(\Bv_0)&:=  \min\left\{\frac{1}{2}\log
\left|\frac{\Sv+\Nv_1}{(\Sv-\Bv_0)+\Nv_1}\right|,\right.\notag\\
&~\quad\left.\frac{1}{2}\log
\left|\frac{\Sv+\Nv_2}{(\Sv-\Bv_0)+\Nv_2}\right|
\right\}\notag\\
f_1(\Bv_1) &:=
\frac{1}{2}\log\left|\frac{\Bv_1+\Nv_1}{\Nv_1}\right|-\frac{1}{2}\log\left|\frac{
\Bv_1+\Nv_2}{\Nv_2}\right|\notag\\
\text{and} \qquad f_2(\Bv_0,\Bv_1) &:= \frac{1}{2}\log\left|\frac{
(\Sv-\Bv_0)+\Nv_2}{\Bv_1+\Nv_2}\right| \notag\\
&~\quad-
\frac{1}{2}\log\left|\frac{(\Sv-\Bv_0)+\Nv_1}{\Bv_1+\Nv_1}\right|.\notag
\end{align}

Let $(\Bv_0^{\star},\Bv_1^{\star})$ be an optimal solution to the
optimization program (\ref{eq:op-A}). By assumption, the matrix power constraint $\Sv$ is strictly positive definite in the canonical model. Thus,
$(\Bv_0^{\star},\Bv_1^{\star})$ must satisfy the following
Karush-Kuhn-Tucker (KKT) conditions:
\begin{align}
&(\beta_1+\lambda_2)[(\Sv-\Bv_0^{\star})+\Nv_1]^{-1} +\beta_2[(\Sv-\Bv_0^{\star})+\Nv_2]^{-1}
  +\Mv_0\notag\\
&\qquad\qquad\qquad=\lambda_2[(\Sv-\Bv_0^{\star})+\Nv_2]^{-1}+\Mv_2 \label{eq:KKT-1}\\
&\quad(\lambda_1+\lambda_2)(\Bv_1^{\star}+\Nv_1)^{-1}+\Mv_1\notag\\
&\qquad\qquad\qquad=(\lambda_1+\lambda_2)(\Bv_1^{\star}+\Nv_2)^{-1}+\Mv_2
\label{eq:KKT-2}\\
&\Mv_0\Bv_0^{\star}=0, ~\Mv_1\Bv_1^{\star}=0,
~\text{and}~\Mv_2(\Sv-\Bv_0^{\star}-\Bv_1^{\star})=0 \label{eq:KKT-4}
\end{align}
where $\Mv_0$, $\Mv_1$ and $\Mv_2$ are positive semidefinite
matrices, and $\beta_k$, $k=1,2$, are nonnegative real scalars such
that $\beta_k>0$ if and only if
\begin{align*}
\frac{1}{2}\log \left|\frac{\Sv+\Nv_k}{(\Sv-\Bv_0^{\star})+\Nv_k}\right| =
R_0^{\dag}.
\end{align*}
It follows that
\begin{align}
(\beta_1 &+\beta_2)R_0^{\dag}+\lambda_1 R_1^{\dag}+ \lambda_2
R_2^{\dag}\notag\\
&=\frac{\beta_1}{2}\log
\left|\frac{\Sv+\Nv_1}{(\Sv-\Bv_0^{\star})+\Nv_1}\right|+\frac{\beta_2}{2}\log
\left|\frac{\Sv+\Nv_2}{(\Sv-\Bv_0^{\star})+\Nv_2}\right|\notag\\
& \quad
+\lambda_1\left(\frac{1}{2}\log\left|\frac{\Bv_1^{\star}+\Nv_1}{\Nv_1}\right|
-\frac{1}{2}\log\left|\frac{ \Bv_1^{\star}+\Nv_2}{\Nv_2}\right|\right)\notag\\
&\quad+\lambda_2\left(\frac{1}{2}\log\left|\frac{(\Sv-\Bv_0^{\star})+\Nv_2}{\Bv_1^{\star}+\Nv_2}\right| \right.\notag\\
&\qquad\left.
-\frac{1}{2}\log\left|\frac{(\Sv-\Bv_0^{\star})+\Nv_1}{\Bv_1^{\star}+\Nv_1}\right|\right)+\rho.
\label{eq:WsumR-1}
\end{align}

Next, we shall find a contradiction to (\ref{eq:WsumR-1}) through the following three steps.

\subsubsection{Split each receiver into two virtual receivers}
Consider the following canonical MIMO Gaussian broadcast channel with four receivers:
\begin{equation}
\begin{array}{rcl}
\Yv_{1a}[m] & = & \Xv[m]+\Zv_{1a}[m]\\
\Yv_{1b}[m] & = & \Xv[m]+\Zv_{1b}[m]\\
\Yv_{2a}[m] & = & \Xv[m]+\Zv_{2a}[m]\\
\Yv_{2b}[m] & = & \Xv[m]+\Zv_{2b}[m]
\label{eq:CH-A2}
\end{array}
\end{equation}
where $\{\Zv_{1a}[m]\}$, $\{\Zv_{1b}[m]\}$, $\{\Zv_{2a}[m]\}$ and $\{\Zv_{2b}[m]\}$ are
i.i.d. additive vector Gaussian noise processes with zero means and covariance matrices
$\Nv_1$, $\Nv_1$, $\Nv_2$ and $\Nv_2$, respectively.

Suppose that the transmitter has three independent messages $W_0$, $W_1$ and $W_2$, where
$W_0$ is intended for both receivers $1b$ and $2b$, $W_1$ is intended for receiver $1a$
but needs to be kept asymptotically perfectly secret from receiver $2b$, and $W_2$ is
intended for receiver $2a$ but needs to be kept asymptotically perfectly secret from
receiver $1b$. Mathematically, for every $\epsilon>0$, we must have
\begin{align}
\frac{1}{n}I(W_1;\Yv_{2b}^n) \leq \epsilon \quad \mbox{and} \quad
\frac{1}{n}I(W_2;\Yv_{1b}^n) \leq \epsilon \label{eq:PS-2}
\end{align}
for sufficiently large block length $n$. Note that receivers $1a$ and $1b$ are statistically identical to receiver 1 in channel
(\ref{eq:CH-A}), so are receivers $2a$ and $2b$ to receiver 2 in channel (\ref{eq:CH-A}).
We thus conclude that the capacity region of channel (\ref{eq:CH-A2}) is the \emph{same}
as that of channel (\ref{eq:CH-A}) under the same matrix power constraint.

\subsubsection{Construct an enhanced channel}
Let $\Nt$ be a real symmetric matrix satisfying
\begin{align}
\Nt&:=\left(\Nv_1^{-1}+\frac{1}{\lambda_1+\lambda_2}\Mv_1\right)^{-1}
\label{eq:def-Nt1}
\end{align}
which implies that $\Nt \preceq \Nv_1$. Since $\Mv_1\Bv_1^{\star}=0$, following \cite[Lemma~11]{WSS-IT06} we have
\begin{align*}
(\lambda_1+\lambda_2)(\Bv_1^{\star}+\Nt)^{-1}&=(\lambda_1+\lambda_2)(\Bv_1^{\star}+\Nv_1)^{-1}+\Mv_1
\end{align*}
and
\begin{align}
|\Bv_1^{\star}+\Nt||{\Nv_1}|&=\left|{\Bv_1^{\star}+\Nv_1}\right||{\Nt}|.
\label{eq:Enh-3}
\end{align}
Following (\ref{eq:KKT-2}), we may also obtain
\begin{align}
(\lambda_1+\lambda_2)(\Bv_1^{\star}+\Nt)^{-1}&=(\lambda_1+\lambda_2)(\Bv_1^{\star}+\Nv_2)^{-1}+\Mv_2
\label{eq:Enh-4}
\end{align}
which implies that $\Nt \preceq \Nv_2$.

Consider the following enhanced aligned MIMO Gaussian broadcast channel
\begin{equation}
\begin{array}{rcl}
\Yt_{1a}[m] & = & \Xv[m]+\Zt_{1a}[m]\\
\Yv_{1b}[m] & = & \Xv[m]+\Zv_{1b}[m]\\
\Yt_{2a}[m] & = & \Xv[m]+\Zt_{2a}[m]\\
\Yv_{2b}[m] & = & \Xv[m]+\Zv_{2b}[m]
\end{array}
\label{eq:CH-A3}
\end{equation}
where $\{\Zt_{1a}[m]\}$, $\{\Zv_{1b}[m]\}$, $\{\Zt_{2a}[m]\}$ and $\{\Zv_{2b}[m]\}$ are
i.i.d. additive vector Gaussian noise processes with zero means and covariance matrices
$\Nt$, $\Nv_1$, $\Nt$ and $\Nv_2$, respectively.

The message set configuration is the same as that for channel (\ref{eq:CH-A2}). Since
$\Nt \preceq \{\Nv_1,\Nv_2\}$, we conclude that the capacity region of channel
(\ref{eq:CH-A3}) is \emph{at least as large} as that of channel (\ref{eq:CH-A2}) under
the same matrix power constraint.

Furthermore, from (\ref{eq:Enh-4}) we have
\begin{align}
[(\Sv-\Bv_0^{\star}) & +\Nt](\Bv_1^{\star}+\Nt)^{-1}\notag\\
&=[(\Sv-\Bv_0^{\star})+\Nv_2](\Bv_1^{\star}+\Nv_2)^{-1} \label{eq:Enh-7}
\end{align}
and hence
\begin{align}
\left|\frac{(\Sv-\Bv_0^{\star})+\Nt}{\Bv_1^{\star}+\Nt}\right|
&=\left|\frac{(\Sv-\Bv_0^{\star})+\Nv_2}{\Bv_1^{\star}+\Nv_2}\right|.
\label{eq:Enh-9}
\end{align}
Combining (\ref{eq:KKT-1}) and (\ref{eq:Enh-4}), we may obtain
\begin{align}
(\lambda_1 &+\lambda_2)[(\Sv-\Bv_0^{\star})+\Nt]^{-1}\notag\\
&= (\lambda_2+\beta_1)[(\Sv-\Bv_0^{\star})+\Nv_1]^{-1}\notag\\
&\quad +(\lambda_1+\beta_2)[(\Sv-\Bv_0^{\star})+\Nv_2]^{-1}
+\Mv_0. \label{eq:Enh-12}
\end{align}
Substituting (\ref{eq:Enh-3}) and (\ref{eq:Enh-9}) into (\ref{eq:WsumR-1}), we
have
\begin{align}
&(\beta_1+\beta_2)R_0^{\dag}+\lambda_1 R_1^{\dag}+ \lambda_2 R_2^{\dag} \notag\\
&=\frac{\beta_1}{2}\log
\left|\frac{\Sv+\Nv_1}{(\Sv-\Bv_0^{\star})+\Nv_1}\right|+\frac{\beta_2}{2}\log
\left|\frac{\Sv+\Nv_2}{(\Sv-\Bv_0^{\star})+\Nv_2}\right|\notag\\
&\quad
+\lambda_1\left(\frac{1}{2}\log\left|\frac{(\Sv-\Bv_0^{\star})+\Nt}{\Nt}\right|
-\frac{1}{2}\log\left|\frac{ (\Sv-\Bv_0^{\star})+\Nv_2}{\Nv_2}\right|\right)\notag\\
&\quad+\lambda_2\left(\frac{1}{2}\log\left|\frac{(\Sv-\Bv_0^{\star})+\Nt}{\Nt}\right|
-\frac{1}{2}\log\left|\frac{(\Sv-\Bv_0^{\star})+\Nv_1}{\Nv_1}\right|\right)\notag\\
&\quad+\rho.
\label{eq:WsumR-2}
\end{align}

\subsubsection{Outer bound the enhanced channel}
Next, we consider a discrete memoryless broadcast channel with four receivers and three independent messages and provide a single-letter outer bound on the capacity region.

\begin{lemma}\label{lemma:DMC2}
Consider a discrete memoryless broadcast channel $p(\yt_{1a}, y_{1b}, \yt_{2a},y_{2b}|x)$ with four receivers and three independent messages $(W_0,W_1,W_2)$: $W_0$ is intended for both receivers $1b$ and $2b$, $W_1$ is intended for receiver $1a$ but needs to be kept asymptotically perfectly secret from receiver $2b$, and $W_2$ is intended for receiver $2a$ but needs to be kept asymptotically perfectly secret from receiver $1b$. Assume that
\begin{align*}
X \rightarrow \widetilde{Y}_{1a} \rightarrow (Y_{1b}, Y_{2b}) \quad \text{and}
\quad  X \rightarrow \widetilde{Y}_{2a} \rightarrow (Y_{1b}, Y_{2b})
\end{align*}
form two Markov chains. Then, any achievable rate triple $(R_0,R_1,R_2)$ must satisfy
\begin{equation}
\begin{array}{rcl}
R_0 & \le & \min[I(U;Y_{1b}),I(U,Y_{2b})]\\
R_1 & \le & I(X;\widetilde{Y}_{1a}|U)-I(X;Y_{2b}|U)\\
R_2 & \le & I(X;\widetilde{Y}_{2a}|U)-I(X;Y_{1b}|U)
\end{array}
\label{eq:degC}
\end{equation}
for some $p(u,x)$, where $U$ is an auxiliary random variable.
\end{lemma}

The proof follows standard information-theoretic argument and is deferred to Appendix~\ref{app:L2}.

Now, we can combine all previous three steps and obtain an upper bound on the weighted sum rate $(\beta_1 +\beta_2)R_0^{\dag}+\lambda_1 R_1^{\dag}+ \lambda_2 R_2^{\dag}$. By assumption, $(R_0^{\dag},R_1^{\dag},R_2^{\dag})$ is an achievable rate triple for channel (\ref{eq:CH-A}). Then, following Lemma~\ref{lemma:DMC2} we have
\begin{align}
(\beta_1 &+\beta_2)R_0^{\dag}+\lambda_1 R_1^{\dag}+ \lambda_2 R_2^{\dag} \notag\\
& \le \frac{\beta_1}{2}\log \left|2\pi e (\Sv+\Nv_1) \right|+
\frac{\beta_2}{2}\log \left|2\pi e (\Sv+\Nv_2) \right|\notag\\
&\quad +\frac{\lambda_1}{2}\log
\left|\frac{\Nv_2}{\Nt}\right|+
\frac{\lambda_2}{2}\log\left|\frac{\Nv_1}{\Nt}\right|+\eta(\lambda_1,\lambda_2)
\label{eq:BSR-1}
\end{align}
where
\begin{align*}
\eta&(\lambda_1,\lambda_2):=\lambda_1 h(\Xv+\Zt_{1a}|U)+ \lambda_2
h(\Xv+\Zt_{2a}|U) \notag\\
&\quad- (\lambda_2+\beta_1)
h(\Xv+\Zv_{1b}|U)-(\lambda_1+\beta_2) h(\Xv+\Zv_{2b}|U).
\end{align*}
Note that $0 \prec \Nt \preceq \{\Nv_1,\Nv_2\}$, $0\prec
\Bv_0^{\star} \preceq \Sv$, and $\Bv_0^{\star}\Mv_0=0$. By
\cite[Corollary~4]{WLSSV-IT09} and (\ref{eq:Enh-12}), we have
\begin{align}
\eta(\lambda_1,\lambda_2) &\le (\lambda_1+\lambda_2)\log\left|2\pi e
(\Sv-\Bv_0^{\star})+\Nt\right|\notag\\
&\quad- (\lambda_2+\beta_1)\log\left|2\pi e
(\Sv-\Bv_0^{\star})+\Nv_1\right|\notag\\
&\quad - (\lambda_1+\beta_2)\log\left|2\pi e (\Sv-\Bv_0^{\star})+\Nv_2\right|.
\label{eq:BSR-2}
\end{align}
Combining  (\ref{eq:BSR-1}) and (\ref{eq:BSR-2}),
we have
\begin{align}
(&\beta_1+\beta_2)R_0^{\dag}+\lambda_1 R_1^{\dag}+ \lambda_2 R_2^{\dag} \notag\\
&\le \frac{\beta_1}{2}\log
\left|\frac{\Sv+\Nv_1}{(\Sv-\Bv_0^{\star})+\Nv_1}\right|+\frac{\beta_2}{2}\log
\left|\frac{\Sv+\Nv_2}{(\Sv-\Bv_0^{\star})+\Nv_2}\right|\notag\\
&\quad
+\lambda_1\left(\frac{1}{2}\log\left|\frac{(\Sv-\Bv_0^{\star})+\Nt}{\Nt}\right|
-\frac{1}{2}\log\left|\frac{ (\Sv-\Bv_0^{\star})+\Nv_2}{\Nv_2}\right|\right)\notag\\
&\quad+\lambda_2\left(\frac{1}{2}\log\left|\frac{(\Sv-\Bv_0^{\star})+\Nt}{\Nt}\right|
-\frac{1}{2}\log\left|\frac{(\Sv-\Bv_0^{\star})+\Nv_1}{\Nv_1}\right|\right)\notag
\end{align}
which is a contradiction to \eqref{eq:WsumR-2} as $\rho>0$. We thus conclude that the rate region \eqref{eq:SCR-A} over all possible $\Bv_0 \succeq 0$, $\Bv_1 \succeq 0$ and $\Bv_0+\Bv_1 \preceq \Sv$ is indeed the capacity region of the canonical MIMO Gaussian broadcast channel \eqref{eq:CH-A}. This completes the proof of Theorem~\ref{thm:GMBC}.

\begin{remark}
Note that in the enhanced channel \eqref{eq:CH-A3}, both legitimate receivers $1a$ and $2a$ have the \emph{same} noise covariance matrices. This fact greatly simplified the capacity analysis of the enhanced channel and is key to the success of the proposed channel enhancement approach. We mention here that the same technique was also used in \cite{Wei-Thesis} to derive the sum-private-\emph{v.s.}-common message capacity region of the MIMO Gaussian broadcast channel.
\end{remark}

\section{Concluding Remarks}
In this paper we have presented two new results on MIMO Gaussian broadcast channels with confidential messages, leading to a more comprehensive understanding of the fundamental limits of MIMO secret communication. 

First, a matrix characterization of the capacity-equivocation region of the MIMO Gaussian wiretap channel has been obtained, generalizing the previous results \cite{KW-IT10a,KW-IT10b,OH-ITS,LS-IT09,BLPS-EURASIP09} which dealt only with the secrecy capacity of the channel. The result has been obtained via an interesting connection between the rate-equivocation setting and simultaneous private-confidential communication over a discrete memoryless wiretap channel, which allows a matrix characterization of the entire capacity-equivocation region based on the existing characterization of secrecy capacity for the MIMO Gaussian wiretap channel. 

Next, the problem of MIMO Gaussian wiretap channels with two receivers and three independent messages, a common message intended for both receivers, and two mutually confidential messages each intended for one of the receivers but needing to be kept asymptotically perfect secure from the other, has been considered. A precise characterization of the capacity region has been obtained via a channel-enhancement argument, which is a natural extension of the channel-enhancement arguments of \cite{LLPS-IT10} and \cite{Wei-Thesis}.

\appendices

\section{Proof of Lemma~\ref{lemma:DMC}}\label{app:L1}

We first prove the achievability part of the lemma by considering a coding scheme that combines superposition coding, random binning, and rate splitting. Fix
$p(u)p(v|u)p(x|v)$. Split the private message $W_p$ into two
independent submessages $W_p'$ and $W_p''$.

\emph{Codebook generation.} Fix $\delta>0$. Randomly and
independently generate $2^{n(R_p'+\delta)}$ codewords of length $n$
according to $p_U^n$. Label each of the codewords as $u^n_{j}$,
where $j$ is the codeword number. We will refer to the codeword
collection $\{u^n_{j}\}_{j}$ as the $U$-codebook.

\begin{figure}
\includegraphics[width=0.9\linewidth,draft=false]{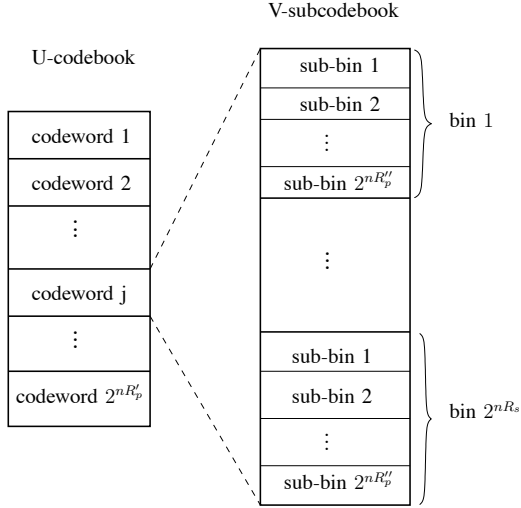}
\caption{Codebook structure.}
\label{fig:nb}
\end{figure}

For each codeword $u^n_{j}$ in the $U$-codebook, randomly and
independently generate $2^{n(R_s+R_p''+T)}$ codewords of length $n$
according to $\prod_{i=1}^n p_{V|U=u_{j}[i]}$. Randomly partition
the codewords into $2^{nR_s}$ bins so that each bin contains
$2^{n(R_p''+T)}$ codewords. Further partition each bin into
$2^{nR_p''}$ sub-bins so that each sub-bin contains $2^{nT}$
codewords. Label each of the codewords as $v^n_{j,k,l,t}$ where $k$
denotes the bin number, $l$ denotes the sub-bin number within each
bin, and $t$ denotes the codeword number within each sub-bin. We
will refer to the codeword collection $\{v^n_{j,k,l,t}\}_{k,l,t}$ as
the $V$-subcodebook corresponding to $u^n_{j}$. Fig.~\ref{fig:nb}
illustrates the overall codebook structure.

\emph{Encoding.} To send a message triple $(w_s,w_p',w_p'')$, the
transmitter first chooses the codeword $u^n_{w_p'}$ from the
$U$-codebook. Next, the transmitter looks into the $V$-subcodebook
corresponding to $u^n_{w_p'}$ and \emph{randomly} (according to a
uniform distribution) chooses a codeword $v^n_{w_p',w_s,w_p'',t}$
from the $w_p''$th sub-bin of the $w_s$th bin. Once a
$v^n_{w_p',w_s,w_p'',t}$ is chosen, an input sequence $x^n$ is
generated according to $\prod_{i=1}^n
p_{X|V=v_{w_p',w_s,w_p'',t}[i]}$ and is then sent through the
channel.

\emph{Decoding at receiver 1.} Given $y_1^n$, receiver 1 looks into
the codebooks $U$ and $V$ and searches for a pair of codewords
$(u^n_{j},v^n_{j,k,l,t})$ that are jointly typical with $y_1^n$. In
the case when
\begin{eqnarray}
R_p'& < & I(U;Y)\label{eq:FM1}\\
\mbox{and} \quad R_s+R_p''+T  & < & I(V;Y|U)\label{eq:FM2}
\end{eqnarray}
with high probability the transmitted codeword pair
$(u^n_{w_p'},v^n_{w_p',w_s,w_p'',t})$ is the only one that is
jointly typical with $y_1^n$.

\emph{Security at receivers 2 and 3.} Fix $\epsilon>0$. In the case
when
\begin{equation}
R_p''+T > I(V;Z|U)\label{eq:FM3}
\end{equation}
we have \cite[Theorem~1]{CK-IT78}
\begin{equation}
\frac{1}{n}I(W_s;Z^n|W_p') \leq \epsilon \label{eq:eqv3}
\end{equation}
for sufficiently large $n$.  Since $W_s$ and $W_p'$ are independent,
we have from \eqref{eq:eqv3} that
\begin{eqnarray}
\frac{1}{n}I(W_s;Z^n) & \leq & \frac{1}{n}I(W_s;Z^n,W_p')\nonumber\\
& = &\frac{1}{n}I(W_s;Z^n|W_p')\nonumber\\ & \leq &
\epsilon\nonumber
\end{eqnarray}
i.e., the message $W_s$ is asymptotically perfectly secure at the
eavesdropper.

To summarize, for any given $p(u)p(v|u)p(x|v)$ and any $T\geq0$, any
rate triple $(R_s,R_p',R_p'')$ that satisfies
\eqref{eq:FM1}--\eqref{eq:FM3} is achievable. Note that
\begin{equation}
R_p=R_p'+R_p''. \label{eq:FM4}
\end{equation}
Eliminating $T$, $R_2'$ and $R_2''$ from
\eqref{eq:FM1}--\eqref{eq:FM3} and \eqref{eq:FM4} using
Fourier-Motzkin elimination, we may conclude that any rate pair
$(R_s,R_p)$ satisfying \eqref{eq:SR-DMC} is achievable.

To prove the converse part of the lemma, we first consider an upper bound on the confidential message rate $R_s$. The perfect secrecy condition (\ref{eq:eqv2}) implies that for every
$\epsilon>0$,
\begin{align}
H(W_s|Z^{n})& \ge H(W_s)-n\epsilon. \label{eq:eqv4}
\end{align}
On the other hand, Fano's inequality \cite[Ch.~2.11]{CT-B91} implies
that for every $\epsilon_0>0$,
\begin{align}
H(W_s,W_p|Y^{n}) &\le \epsilon_0 \log\left[2^{n(R_s+R_p)}-1\right]
            +h(\epsilon_0) \notag\\
&:= n\delta.  \label{eq:sd1}
\end{align}
Applying (\ref{eq:eqv4}) and (\ref{eq:sd1}), we have
\begin{align}
nR_s & = H(W_s) \notag \\
           &\le  \bigl[H(W_s|Z^{n})+n \epsilon\bigr]+ \bigl[n \delta-H(W_s,W_p|Y^{n})\bigr]\notag\\
           &\le  H(W_s,W_p|Z^{n})-H(W_s,W_p|Y^{n}) + n (\epsilon+\delta).  \label{eq:apout2}
\end{align}
By the chain rule of the mutual information \cite[Ch.~2.5]{CT-B91},
\begin{align}
n(R_s -\epsilon-\delta)&\le I(W_s,W_p;Y^{n})-I(W_s,W_p;Z^{n})\notag\\
              & =  \sum_{i=1}^{n}\bigl[ I(W_s,W_p;Y_i|Y^{i-1})\notag\\
              &\qquad \quad -I(W_s,W_p;Z_i|Z_{i+1}^{n})\bigr] \notag\\
& =  \sum_{i=1}^{n}\bigl[ I(W_s,W_p;Y_i|Y^{i-1},Z_{i+1}^{n})\notag\\
              &\qquad \quad -I(W_s,W_p;Z_i|Y^{i-1},Z_{i+1}^{n})\bigr] \label{eq:apout3}
\end{align}
where the last equality follows from \cite[Lemma~7]{CK-IT78}. Let
\begin{align}
U_i&:= \left(Y^{i-1},Z_{i+1}^{n}\right) \quad \mbox{and} \quad
V_i:=\left(W_s,W_p,U_i\right) \label{eq:def-Ui}
\end{align}
and we have from (\ref{eq:apout3}) that
\begin{align}
n(R_s-\epsilon-\delta)&\le \sum_{i=1}^n\left[I(V_i;Y_i|U_i)-I(V_i;Z_i|U_i)\right].
\label{eq:apout4}
\end{align}

Next, we consider an upper bound on the sum private-confidential message rate $R_s+R_p$. By
(\ref{eq:sd1}),
\begin{align}
n(R_s+R_p) & = H(W_s,W_p) \notag \\
     &\le I(W_s,W_p; Y^{n})-n\delta.
\end{align}
Applying the chain rule of the mutual information
\cite[Ch.~2.5]{CT-B91}, we have
\begin{align}
n(R_s+R_p-\delta) & \le \sum_{i=1}^{n} I(W_s,W_p; Y_i|Y^{i-1})\notag\\
& \le \sum_{i=1}^{n} I(W_s,W_p,Y^{i-1},Z_{i+1}^n; Y_i) \notag\\
&=\sum_{i=1}^{n} I(V_i; Y_i).
\label{eq:apout5}
\end{align}
Applying the standard single-letterization procedure (e.g., see
\cite[Ch.~14.3]{CT-B91}) to \eqref{eq:apout4} and \eqref{eq:apout5}, we have the desired converse result for
Lemma~\ref{lemma:DMC}.

\section{Proof of Lemma~\ref{lemma:DMC2}}\label{app:L2}

The perfect secrecy condition (\ref{eq:PS-2}) implies that for every $\epsilon>0$,
\begin{subequations} \label{eq:eeqv}
\begin{align}
&          & H(W_1|Y_{2b}^{n})& \ge H(W_1)-n\epsilon & \label{eq:eeqv1} \\
&\text{and}& H(W_2|Y_{1b}^{n})& \ge H(W_2)-n\epsilon. & \label{eq:eeqv2}
\end{align}
\end{subequations}
On the other hand, Fano's inequality \cite[Chapter~2.11]{CT-B91} implies that for every
$\epsilon_0>0$,
\begin{subequations} \label{eq:ssd}
\begin{align}
\max[ H&(W_0|Y_{1b}^{n}),\;H(W_0|Y_{2b}^{n})]\notag\\
             &\le \epsilon_0 \log\left(2^{nR_0}-1\right)+h(\epsilon_0) := n\delta_0  \label{eq:ssd0}\\
H&(W_1|\widetilde{Y}_{1a}^{n}) \notag\\
&\le \epsilon_0 \log\left(2^{nR_1}-1\right)
            +h(\epsilon_0) := n\delta_1 \label{eq:ssd1}\\
\text{and} \qquad H&(W_2|\widetilde{Y}_{2a}^{n}) \notag\\
&\le \epsilon_0 \log\left(2^{nR_2}-1\right)
            +h(\epsilon_0) := n\delta_2. \label{eq:ssd2}
\end{align}
\end{subequations}
Let 
\begin{align}
U_i:=(W_0, Y_{1b}^{i-1},Y_{2b,i+1}^{n}) \label{eq:def-U}
\end{align}
which satisfies the Markov chain
\begin{align}
U_i \rightarrow X_i \rightarrow (\widetilde{Y}_{1a}, \widetilde{Y}_{2a}, Y_{1b}, Y_{2b}).
\end{align}
We first bound $R_0$ based on (\ref{eq:ssd0}) as follows:
\begin{align}
nR_0 & = H(W_0) \notag \\
    &\le I(W_0; Y_{1b}^{n})+n\delta_0 \notag\\
    &=   \sum_{i=1}^{n}I(W_0; Y_{1b,i}|Y_{1b}^{i-1})+n\delta_0 \notag\\
    &\le \sum_{i=1}^{n}I(W_0,Y_{1b}^{i-1},Y_{2b,i+1}^{n}; Y_{1b,i})+n\delta_0 \notag\\
    &=   \sum_{i=1}^{n}I(U_i; Y_{1b,i})+n\delta_0. \label{eq:UPR0-1}
\end{align}
Similarly, we have
\begin{align}
nR_0  &\le I(W_0; Y_{2b}^{n})+n\delta_0 \notag\\
    &=   \sum_{i=1}^{n}I(W_0; Y_{2b,i}|Y_{2b,i+1}^{n})+n\delta_0 \notag\\
    &\le \sum_{i=1}^{n}I(W_0,Y_{1b}^{i-1},Y_{2b,i+1}^{n}; Y_{2b,i})+n\delta_0 \notag\\
    & =  \sum_{i=1}^{n}I(U_i; Y_{2b,i})+n\delta_0. \label{eq:UPR0-2}
\end{align}
Next, we bound $R_1$ based on (\ref{eq:eeqv1}) and (\ref{eq:ssd1}) as follows:
\begin{align}
nR_1 & = H(W_1) \notag \\
    &\le \bigl[H(W_1|Y_{2b}^{n})+n \epsilon\bigr]+ \bigl[n \delta_1-H(W_1|\widetilde{Y}_{1a}^{n})\bigr]\notag\\
    & =  H(W_1|W_0,Y_{2b}^{n})+I(W_1;W_0|Y_{2b}^{n})-H(W_1|\widetilde{Y}_{1a}^{n}) \notag\\
    &\quad +n (\epsilon+ \delta_1)\notag\\
    & \le  H(W_1|W_0,Y_{2b}^{n})+H(W_0|Y_{2b}^{n})-H(W_1|W_0,\widetilde{Y}_{1a}^{n}) \notag\\
    &\quad +n (\epsilon+ \delta_1). \label{eq:app-r1b1}
\end{align}
Substituting (\ref{eq:ssd1}) into (\ref{eq:app-r1b1}), we may obtain
\begin{align}
nR_1&\le H(W_1|W_0,Y_{2b}^{n})-H(W_1|W_0,\widetilde{Y}_{1a}^{n}) \notag\\
    &\quad + n (\epsilon+\delta_0+\delta_1)\notag \\
    & =  I(W_1;\widetilde{Y}_{1a}^{n}|W_0)-I(W_1;Y_{2b}^{n}|W_0) \notag\\
    &\quad + n (\epsilon+\delta_0+\delta_1). \label{eq:app-r1b2}
\end{align}
Applying \cite[Lemma~7]{CK-IT78}, (\ref{eq:app-r1b2}) can be rewritten as
\begin{align}
nR_1 & \le  \sum_{i=1}^{n}
\bigl[I(W_1;\widetilde{Y}_{1a,i}|W_0,\widetilde{Y}_{1a}^{i-1},Y_{2b,i+1}^{n})\notag\\
 & \quad
    -I(W_1;Y_{2b,i}|W_0,\widetilde{Y}_{1a}^{i-1},Y_{2b,i+1}^{n})\bigr]+ n (\epsilon+\delta_0+\delta_1)\notag \\
    &\le \sum_{i=1}^{n}
    \bigl[I(X_i;\widetilde{Y}_{1a,i}|W_0,\widetilde{Y}_{1a}^{i-1},Y_{2b,i+1}^{n}) \notag\\
 & \quad -I(X_i;Y_{2b,i}|W_0,\widetilde{Y}_{1a}^{i-1},Y_{2b,i+1}^{n})\bigr] + n(\epsilon+\delta_0+\delta_1)     \label{eq:UPR1-2}
\end{align}
where (\ref{eq:UPR1-2}) follows from the Markov chain
$$W_1 \rightarrow X_i \rightarrow \widetilde{Y}_{1a,i} \rightarrow Y_{2b,i}.$$
Moreover, due to the Markov chain
\begin{align}
(W_0,\widetilde{Y}_{1a,i},Y_{2b,i+1}^{n})\rightarrow \widetilde{Y}_{1a}^{i-1}\rightarrow Y_{1b}^{i-1}
\end{align}
we can further bound $R_1$ as
\begin{align}
nR_1 &\le \sum_{i=1}^{n}
    \bigl[I(X_i;\widetilde{Y}_{1a,i}|W_0,\widetilde{Y}_{1a}^{i-1},Y_{1b}^{i-1},Y_{2b,i+1}^{n})\notag\\
    &\quad -I(X_i;Y_{2b,i}|W_0,\widetilde{Y}_{1a}^{i-1},Y_{1b}^{i-1},Y_{2b,i+1}^{n})\bigr] \notag\\
    &\quad + n (\epsilon+\delta_0+\delta_1)\notag\\
    &= \sum_{i=1}^{n}\bigl[I(X_i;\widetilde{Y}_{1a,i}|U_i,\widetilde{Y}_{1a}^{i-1}) \notag\\
    &\quad -I(X_i;Y_{2b,i}|U_i,\widetilde{Y}_{1a}^{i-1})\bigr] + n (\epsilon+\delta_0+\delta_1)
   \label{eq:UPR1-3}\\
   &= \sum_{i=1}^{n} \bigl[I(X_i;\widetilde{Y}_{1a,i}|U_i)
    -I(X_i;Y_{2b,i}|U_i)\bigr] \notag\\
    &\quad -\bigl[I(\widetilde{Y}_{1a}^{i-1};\widetilde{Y}_{1a,i}|U_i)
    -I(\widetilde{Y}_{1a}^{i-1};Y_{2b,i}|U_i)\bigr] \notag\\
    &\quad + n (\epsilon+\delta_0+\delta_1)\notag\\
   &\le \sum_{i=1}^{n}
    \bigl[I(X_i;\widetilde{Y}_{1a,i}|U_i)
    -I(X_i;Y_{2b,i}|U_i)\bigr] \notag\\
    &\quad   + n (\epsilon+\delta_0+\delta_1)
   \label{eq:UPR1-4}
\end{align}
where (\ref{eq:UPR1-3}) follows from the definition of $U_i$ in (\ref{eq:def-U}), and
(\ref{eq:UPR1-4}) follows from the fact that $Y_{2b,i}$ is degraded with respect to 
$\widetilde{Y}_{1a,i}$ so $I(\widetilde{Y}_{1a}^{i-1};Y_{2b,i}|U_i) \leq I(\widetilde{Y}_{1a}^{i-1};\widetilde{Y}_{1a,i}|U_i)$.

Following the same steps as those in (\ref{eq:app-r1b1})--(\ref{eq:UPR1-4}), we may obtain
\begin{align}
nR_2 &\le \sum_{i=1}^{n} \bigl[I(X_i;\widetilde{Y}_{2a,i}|U_i) \notag\\
    &\quad -I(X_i;Y_{1b,i}|U_i)\bigr] + n (\epsilon+\delta_0+\delta_2).
   \label{eq:UPR2-1}
\end{align}

Finally, applying the standard single-letterization procedure (e.g., see
\cite[Chapter~14.3]{CT-B91}) to (\ref{eq:UPR0-1}), (\ref{eq:UPR0-2}), (\ref{eq:UPR1-4})
and (\ref{eq:UPR2-1}) proves the desired result \eqref{eq:degC} for Lemma~\ref{lemma:DMC2}.

\end{document}